\documentclass[acus]{JAC2001}

\usepackage{graphicx}

\begin{document}
\title{Ultra High Precision with a Muon Storage Ring$^{\dag}$ 
}
 
\author{B. Lee Roberts\thanks{roberts@bu.edu,
 $^{\dag}$Invited talk at the European 
Particle Accelerator Conference, June 2002},
Department of Physics, Boston University, Boston, MA 02215, USA \\
for the Muon $(g-2)$ Collaboration\cite{colab}
}
\maketitle

\begin{abstract}
The Muon  $(g-2)$ experiment, E821, at the Brookhaven AGS has the goal to measure
the muon anomalous magnetic moment to a relative accuracy of 
$\pm 3.5 \times 10^{-7}$.  A superferric 14 m diameter storage ring has been
constructed and an averaged
magnetic field uniformity over the 90 mm diameter
muon storage region of $\pm$ 1 part per million
(ppm) has been achieved. A truncated double-cosine superconducting septum
magnet (the inflector) was constructed along with a fast non-ferric 
kicker.  The performance of the storage ring, along with the physics results
are reviewed.
\end{abstract}

\section{Introduction}

The measurement of magnetic moments has been
important in advancing our knowledge
of sub-atomic physics since the famous 1921 paper of Stern,\cite{stern}
 which laid out the principles of what we now call the ``Stern-Gerlach
experiment''.  The experimental and theoretical developments in the
study of the electron's magnetic moment represent one of the great success
stories of modern physics, with the experiment reaching a relative accuracy
of $\sim 4$ parts in $10^9$ (parts per billion)\cite{vd} and the theory being
constrained by our knowledge of the fine-structure constant $\alpha$, rather
than by the 
eight-order and tenth-order QED calculations.\cite{kinalpha} 

The gyromagnetic ratio $g$ is defined by 
$\vec \mu_s = g ( {e / 2m} ) \vec s$, 
where $\vec s$ is the spin angular momentum, and $\vec \mu$ is the
magnetic moment resulting from this angular momentum.
The Dirac equation predicts that $g\equiv 2$, but radiative corrections
increase the value at the part per mil level.
The Particle Data Tables  define the magnetic moment
as $\mu = (1+a) {e \hbar / 2 m}$ where  $a = (g - 2)/ 2$
is the anomalous magnetic moment (or simply the anomaly). 

When E821 began in the early 1980s, 
$a_{\mu}$ was known to 7.3 parts per million (ppm).\cite{cern3}  
The E821 Collaboration
has reported three new measurements with relative accuracies of 13, 5 and
1.3 ppm respectively.\cite{carey, brown1, brown2}

To the level of the experimental accuracy, the electron anomaly can be
described by the QED of $e^{\pm}$ and photons, with the contribution of
heavier virtual particles entering at a level below 4 ppb.  The larger mass of
the muon permits heavier virtual particles to contribute, and the enhancement
factor is $ {\sim  ( {m_{\mu} /  m_e} )^2} \sim 40,000$.  The CERN
measurement observed the effect on $a_{\mu}$  of virtual hadrons 
at the 10 standard deviation level.\cite{cern3}
The standard model value of $a_{\mu}$ consists of 
QED, strong interaction and  weak radiative corrections,
and a significant 
deviation from the calculated standard model value would represent
a signal for non-standard model physics.

\section{The Experimental Technique}

The method used in the third CERN experiment and the BNL experiment are
very similar, save the use of direct muon injection into the storage ring
which was developed by the E821 collaboration.  They are based on the
fact that for $g\neq 2$ (or more precisely $a_{\mu} > 0$) the spin gets
ahead of the momentum vector when a muon travels transversely to a 
magnetic field.  The Larmor and Thomas spin-precession and the momentum 
precession frequencies are
\begin{equation}
 \omega_S = {geB \over 2 m c} + (1-\gamma) {e B \over \gamma mc};\qquad
 \omega_C = {e B \over mc \gamma}
\end{equation}
and the difference frequency gives the frequency with which the spin
precesses relative to the momentum, 
\begin{equation}
 \qquad
\omega_a = \omega_S - \omega_C = ({g-2 \over 2}) {eB \over mc}
\label{eq:omeganoE}
\end{equation}
which is proportional to the anomaly, rather than to the full magnetic
moment. A precision measurement of $a_{\mu}$ requires precision measurements
of the precession frequency $\omega_a$  and the magnetic field.
The muon frequency can be measured as accurately as the counting
statistics and detector apparatus permit.  
The design goal for the NMR magnetometer and calibration system
was a field accuracy of about 0.1 ppm.  The $B$ which enters in 
Eq. \ref{eq:omeganoE} is the average field seen by the ensemble of muons
in the storage ring, 
$<B>_{\phi} = <\int M(r,\theta)B(r, \theta) rdr d\theta>_{\phi}$
where $\phi$ is the azimuthal angle around the ring, $r,\theta$ are the
coordinates at a single slice of azimuth centered at the middle of the
90 mm diameter muon storage region.  $M(r,\theta)$ is the moment (multipole)
distribution of the muon distribution, and couples multipole by multipole
with the magnetic field.  It is very difficult to obtain adequate information
on the higher
moments of the muon distribution in the storage ring, so the presence
of higher multipoles in the magnetic field is undesirable.

The need for vertical focusing implies that a gradient field is needed,
but the usual magnetic gradient used in storage rings is ruled out in
our case.  A sufficient magnetic gradient for vertical focusing would
spoil the ability to use NMR to 
measure the magnetic field to the necessary accuracy, and would
also require
detailed knowledge of the muon distribution.

An electric quadrupole is used instead, taking advantage of the 
``magic''~$\gamma=29.3$ at which an electric field does not contribute to
the spin motion relative to the momentum.  This can be understood from the
famous Thomas-BMT equation
\begin{equation}
\vec \omega_a = {e \over mc}
\left[ a_{\mu} \vec B -
\left( a_{\mu}- {1 \over \gamma^2 - 1}\right) \vec \beta \times \vec E
\right],
\label{eq:tbmt}
\end{equation}
which reduces to Eq. \ref{eq:omeganoE} in the absence of an electric field.
Note that for muons with $\gamma = 29.3$ in an electric field alone,
the spin would follow the momentum vector.

The arrangement of a magnetic dipole field combined with an electric
quadrupole field is called a Penning trap in atomic physics.
However with a 14 m diameter and $\sim700$ T weight, 
the scale of our trap is
quite different from the usual one.\cite{vd}  

In order to meet the conditions discussed above, a goal of $\pm 1$ ppm
uniformity of the $<B>$-field over the storage region was set and met.
A round beam profile was chosen, since sharp corners
would imply large higher moments for $M(r,\theta)$.  
Given the projected knowledge of the
muon distribution, the allowable strength of the 
quadrupole and higher magnetic
multipoles was also determined.  The quadrupoles
are arranged in a four-fold symmetry covering 43\% of the
ring.  This geometry has the advantage that $\beta_{\rm max} \simeq 
\beta_{\rm min}$ to about 5\% so the beam does not breathe very much,
making the average field calculation easier.

In our optimal running conditions, the 24 GeV/c proton beam in the AGS is
accelerated in  12 proton bunches, each with an intensity of
about $5 \times 10^{12}$ protons.  The bunches are extracted one at a time
at 33 ms intervals,
and brought down a transport line to a Ni production target. The time
distribution of the proton beam has $\sigma_t \simeq 25$ ns.  The AGS
cycle time is about 2.8 s.

 Pions at
$0^{\circ}$ are momentum analyzed and then brought into a 72 m straight
FODO decay channel where muons are born.  A second momentum slit permits one
to choose forward muons about 1.6\% below the pion momentum.  These forward
muons have $>90$\% longitudinal polarization, and are injected into
the ring where their spin precesses.  We store on the order of $10^4$ muons
per fill of the ring.

\begin{figure}[htb]
\centering
\includegraphics*[width=75mm]{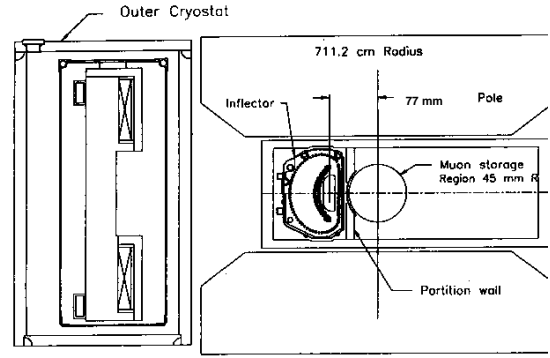}
\caption{ The inflector exit-vacuum chamber geometry. The center of the 
storage ring is to the right.  The gap between the pole pieces is 180~mm,
and the inflector exit is 18 $\times$ 56~mm$^2$ (ignoring the chamfer on the 
outer radius corners).
\label{fg:infgeo2}}
\end{figure}

Unlike the conventional storage ring made of lumped elements, the requirement
of a field uniform at the ppm level precluded breaks in the storage ring
magnet. Thus the beam must be brought in through the fringe field of the
storage ring to a point close the the central orbit.
This is achieved by the use of a 1.7 m 
superconducting inflector magnet\cite{meng}
which nulls the field where the beam enters, but does not leak flux into the
storage region, spoiling the precision field. An elevation view of the 
inflector exit, and the magnet is shown in Fig. \ref{fg:infgeo2}.
This unconventional set of conditions means that there is no phase-space
matching between the incoming beam and the storage ring.  This mismatch
reduces the calculated injection efficiency to $\sim 8.7$\%.  

A kick of about 0.1 Tm is needed to bring the beam onto a stable orbit.
This is achieved with three 1.7 m long ferrite-free kickers,\cite{kicker} 
which can be
thought of as single-loop pulsed magnets carrying a current of 
4,200 A.  The minimum inductance achievable of 1.6 $\mu$H 
limited the peak current to 4200A, and resulted in a current-pulse 
width $\sim2.5$ times greater
than the cyclotron period.  This less than optimal kicker pulse reduces the
injection efficiency to about 7.3\%.  Nevertheless the number of stored muons
per fill is almost a factor of 100 over that available in the final CERN
experiment,\cite{cern3} and the injection-related background seen by the
detectors is down by a factor of 50.\cite{carey,brown1}

\begin{figure}[htb]
\centering
  \includegraphics*[width=59mm]{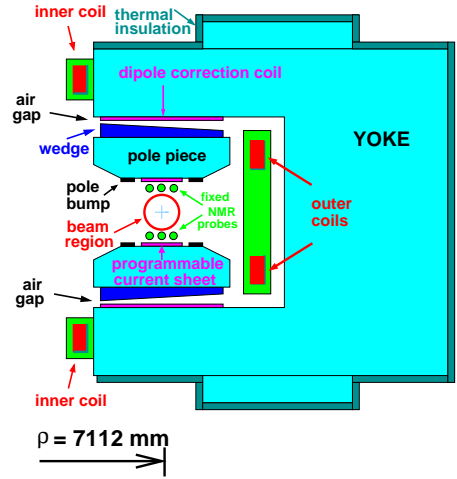}
  \caption{ A schematic of the magnet cross section.}
       \label{fg:magcross}
\end{figure}

The precision magnetic field is at the heart of the experiment.\cite{danby}
  A profile of
the storage ring magnet is given in Fig. \ref{fg:magcross}.

The air-gap dominates the reluctance of the magnetic circuit, and the
pole-pieces are made of very low carbon, high quality magnet steel.  The
angle of the iron wedge in the air gap is used to eliminate the quadrupole
inherent in a ``C'' magnet, and the wedge can be moved radially to adjust the
dipole locally.  The pole bumps were ground individually to minimize the
sextupole component, and the pole face windings (programmable current sheet)
permit one to minimize the higher multipoles (on average), 
which do not vary much around
the ring.

\begin{figure}[htb]
\centering
  \includegraphics*[width=75mm]{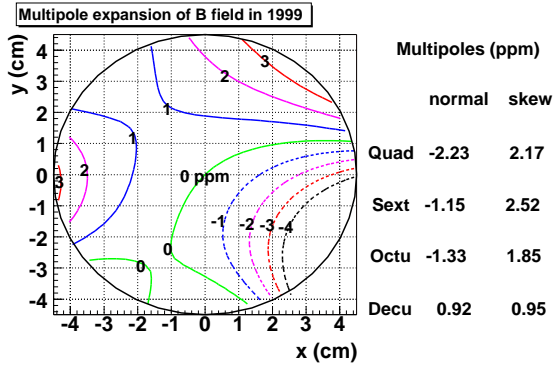}
  \caption{Contour map of the magnetic field averaged over azimuth for 1999.
The contours show 1 ppm differences.}
       \label{fg:field1}
\end{figure}

\begin{figure}[htb]
\centering
  \includegraphics*[width=60mm]{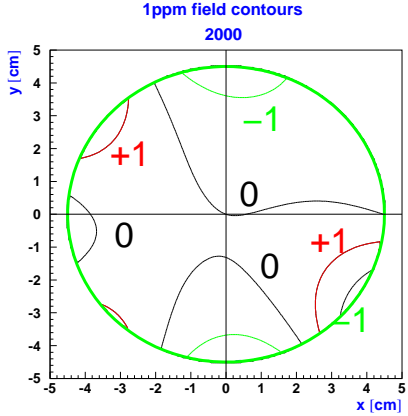}
  \caption{Contour map of the magnetic field averaged over azimuth for 2000.
The contours show 1 ppm differences.}
       \label{fg:field2}
\end{figure}

The success in shimming the magnet can be seen in Figs. \ref{fg:field1} and
\ref{fg:field2}, which show the average field from 1999 and 2000.  The 
poorer field quality in '99 came from a damaged passive
superconducting shield around the inflector, which permitted flux leakage into
the storage ring.  In 2000 the damaged inflector was replaced, and
we met the goal of $\pm 1$ ppm uniformity.
Several important parameters of the storage ring are given in Table 
\ref{tb:ring}.

The experimental signal is the $e^{\pm}$ from $\mu^{\pm}$ decay.  
Muon decay is a three-body decay, so the 3.1 GeV muons produce a continuum
of positrons (electrons) from the endpoint energy down.  Since the highest
energy  $e^{\pm}$ are correlated with the muon spin, if one counts high energy 
 $e^{\pm}$ as a function of time, one gets an exponential from muon decay
modulated by the $(g-2)$ precession. The expected form for the positron time
spectrum is 
\begin{equation}
f(t) =  {N_0} e^{- \lambda t } 
[ 1 + {A} \cos ({\omega_a} t + {\phi})] 
\label{eq:5pm}
\end{equation}

\begin{figure}[htb]
\centering
  \includegraphics*[width=55mm]{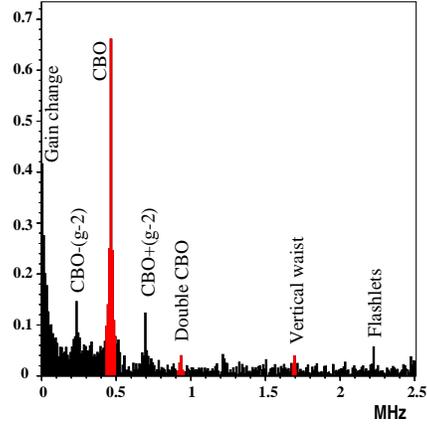}
  \caption{A Fourier Transform of the fit residuals from a 5-parameter fit to
the 1999 data set.
       \label{fg:FT5pm}}
\end{figure}

However, a Fourier analysis of the residuals from this five parameter fit
shows a number of frequency components which can only be understood after a
discussion of the beam dynamics in the ring (see  Fig. \ref{fg:FT5pm}).

\begin{table}[hbt]
\begin{center}
\caption{Parameters of the storage ring.}
\begin{tabular}{|l|l|} \hline
{\it Parameter} & {\it Value} \\
\hline
$(g-2)$ Frequency & $f_a \sim 0.23 \times 10^6$ Hz \\
$(g-2)$ Period & $\tau_a =  4.37 \mu$s \\
\hline
{Muon Kinematics} & $p_{\mu}=3.094$ GeV/c \\
 &  $\gamma_{\mu}=29.3$ \\
 & $ \gamma \tau = 64.4\ \mu$s \\
\hline
Cyclotron Period & $ \tau_{cyc} = 149$ ns  \\
\hline
Central Radius & $\rho = 7112$ mm  \\
\hline
Magnetic Field &$ B_0 = 1.451$ T\\
\hline
 Storage Aperture & $9.0$ cm circle \\
\hline
In one lifetime:  & 432 revolutions  \\
&\ 14.7 {\it (g-2)} periods  \\
\hline
\end{tabular}
\label{tb:ring}
\end{center}
\end{table}

\section{Beam Dynamics}

The $(g-2)$ ring is a weak focusing ring with  
\begin{equation}
n = {\kappa R_0 \over \beta B_0}
\label{eq:n}
\end{equation}
where $ \kappa$ is the electric quadrupole  gradient.  Several $n$~-~values
were used for data acquisition: 
$n = 0.137,\ 0.142$ and 0.122.
The horizontal and 
vertical betatron frequencies are approximately given by
\begin{equation}
 f_x = f_C \sqrt{1-n}\simeq 0.929 f_C \quad
f_y = f_C \sqrt{n} \simeq 0.37 f_C
\label{eq:betafreq}
\end{equation}
where $f_C$ is the cyclotron frequency and the numerical values 
assume $n=0.137$.

The detector acceptance depends on the radial position of the muon
when it decays, so any coherent radial beam
motion will amplitude modulate the decay $e^{\pm}$ distribution.
The principal frequency will be the
 ``Coherent Betatron Frequency''
\begin{equation}
f_{\rm CBO} = f_C - f_x = (1 - \sqrt{1-n})f_C 
\label{eq:cbo}
\end{equation}
which is the frequency a single fixed detector sees the beam moving
 coherently back and forth.  An alternate way of thinking about this
frequency is to view the ring as a spectrometer where the inflector exit is
imaged at each successive betatron wavelength.  In principle an inverted
 image appears at half a betatron wavelength, but the radial image is spoiled
 by the $\pm 0.5$\% momentum dispersion of the ring.  A given
 detector will see the beam move radially with the CBO frequency, which is
 also the frequency that the horizontal waist precesses around the ring.
  However, since there is no dispersion in the vertical dimension, the vertical
waist is reformed every half a wavelength.
  The CBO frequency and its
sidebands are clearly visible in the Fourier spectrum, and the vertical
waist is just seen.
A number of frequencies in the ring are tabulated in Table~\ref{tb:freq}
\begin{table}[hbt]
\begin{center}
\caption{Frequencies in the $(g-2)$ storage ring for $n~=~0.137$.}
\begin{tabular}{|l|l|l|l|} \hline
{\it Quantity } & {\it Expression} &{\it Frequency } & {\it Period } 
 \\
 & & &  \\
\hline
$f_a$ &  ${e \over 2 \pi mc} a_{\mu}  B$ & 0.23 MHz & 4.37 $\mu$s  \\
\hline
$f_c$ & ${v \over 2 \pi R_0}$ & 6.7 MHz & 149 ns \\
\hline 
$f_x$ & $\sqrt{1-n}f_c$ & 6.23 MHz & 160 ns \\
$f_y$ & $\sqrt{n}f_c$ & 2.48 MHz & 402 ns \\
\hline
$f_{\rm CBO}$ & $f_c - f_x$ & 0.477 MHz & 2.10  $\mu$s \\
$f_{\rm VW}$  & $f_c - 2f_y$ & 1.74 MHz & 0.574  $\mu$s \\
\hline
\end{tabular}
\label{tb:freq}
\end{center}
\end{table}

\begin{figure}[htb]
\centering
  \includegraphics*[width=55mm]{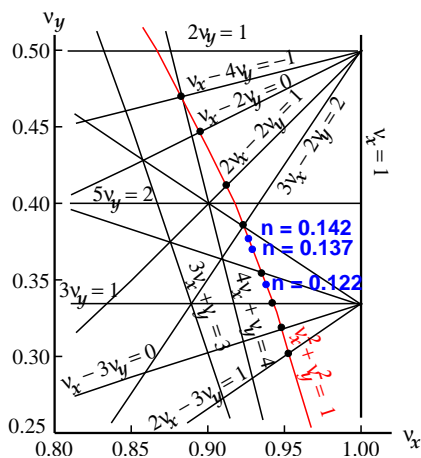}
  \caption{The tune plane showing the three operating points used 
during our three year's of running.
       \label{fg:tunepl}}
\end{figure}

The tune plane is shown in Fig.\ref{fg:tunepl} which shows resonance lines up
to fifth order.  Of the three $n$-values used for data collection, the
$n=0.137$ tune had a CBO frequency uncomfortably close to the second harmonic
of $(g-2)$ (see Table \ref{tb:freq} putting an AM sideband just by
the  $(g-2)$ frequency.  This nearby sideband
and has forced us to to work very hard to
understand the CBO and how its related phenomena affect the value of 
$\omega_a$ from the fits to the data. The 2001 data set was taken with
the other two tunes, which substantially reduced our sensitivity to the CBO.

\section{Monitoring the Beam Profile}

Three tools were available to us to monitor the muon distribution.  By
studying the beam debunching after injection, one can gain information on the
distribution of equilibrium radii in the storage ring.  A wire chamber system
located at one position around the ring permitted us to measure the
trajectories of the $e^{\pm}$ and by tracing the trajectory back to the
point where it is tangent to the ring, reproduces the decay position to
within a few mm.  The ring was equipped with two sets of scintillating fiber
beam monitors which could be plunged into the storage region. Each set
consisted of an $x$ and $y$ plane of seven 0.5 mm diameter scintillating
fibers which covered the beam region. 

\begin{figure}[htb]
\centering
  \includegraphics*[width=70mm]{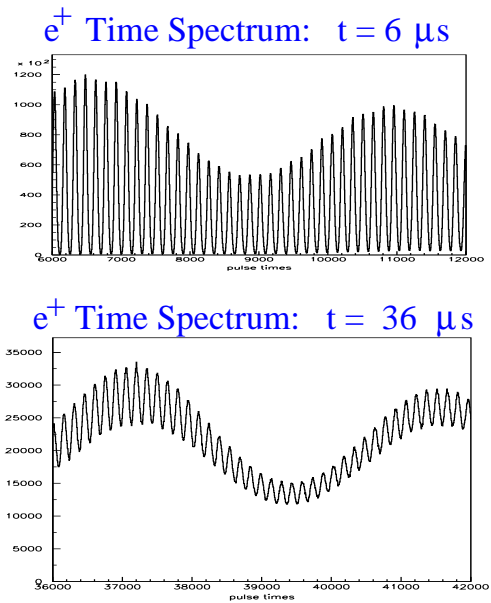}
  \caption{The time spectrum of a single detector soon after injection.
The spikes are separated by the cyclotron period of 149 ns.
       \label{fg:fastrot}}
\end{figure}

\begin{figure}[htb]
\centering
  \includegraphics*[width=50mm]{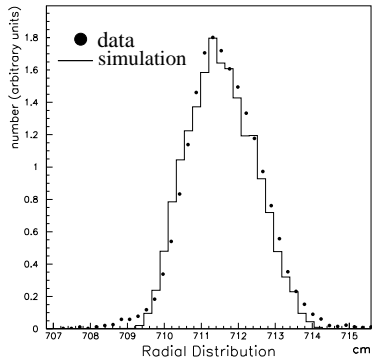}
  \caption{The distribution of equilibrium radii obtained from the 
beam de-bunching.
       \label{fg:rdist}}
\end{figure}

In Fig. \ref{fg:fastrot} the signal from a single detector is shown at two
different times following injection.  The bunched beam is seen very clearly
in the top figure, with the 149 ns cyclotron period being obvious.  The slow
amplitude modulation comes from the $(g-2)$ precession.  By 36 $\mu$s the
beam has largely de-bunched.  In Fig. \ref{fg:rdist} the inferred
distribution
of equilibrium radii is shown along with that obtained from a monte carlo
tracking code.   The agreement is seen to be good, and the measured
distribution was used in determining the average magnetic field and the
radial electric field correction.

\begin{figure}[htb]
\centering
  \includegraphics*[width=50mm]{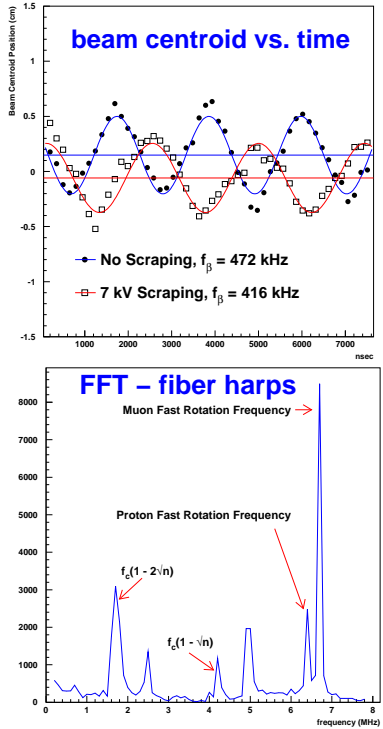}
  \caption{The beam centroid motion and its Fourier transform.
       \label{fg:fbm}}
\end{figure}

The scintillating fiber monitors showed clearly the vertical and horizontal
tunes as expected.  Fig. \ref{fg:fbm} the beam centroid motion is
shown, both with the quadrupoles powered asymmetrically during scraping, and
symmetrically after scraping.  A Fourier transform of the latter signal shows
clearly the expected frequencies.  The traceback system also clearly sees the
CBO motion.
Additional details on beam dynamics will be available in Ref. \cite{beam}

A physics result with a total error of 1.3 ppm was obtained from our 1999
data set.\cite{brown2}  The result was originally 2.6 standard deviations
above the predicted standard model value, creating an outpouring of interest,
and much speculation that the first evidence for supersymmetry had been
obtained. However, a sign mistake in a small piece of the theoretical hadronic
contribution improved the agreement to 1.6 $\sigma$. We have an additional 
data set with $\mu^+$ which will have a total error of around 0.8 ppm, and
also a $\mu^-$ data set which should also have a similar error, although the
distribution between statistical and systematic errors is different for the
two data sets.

The theory value is also being refined further, and we hope that both theory
and experiment will have presented new values by September 2002.

\section{Conclusions}

The muon $(g-2)$ collaboration has successfully built and operated a
superferric storage ring with unprecedented field uniformity over such a
large volume.  All aspects of the apparatus, inflector, main magnet, kicker,
electrostatic quadrupoles,
detector system and electronics have achieved the design specifications.
Interest in our new results remains high, and we look forward to improvements
from our theoretical colleagues, which will improve the sensitivity of the
comparison to the standard model.

{\it Acknowledgements} I wish to thank my colleagues, K. Jungmann,
J. Paley, E. Sichtermann, and Y.K. Semertzidis for their critical readings
of,
and suggested improvements for 
this manuscript.

\end{document}